\documentclass{ws-procs9x6-cpt22}
\begin{document}

\newcommand{\refeq}[1]{(\ref{#1})}
\def\etal {{\it et al.}}

\title{On the Hamiltonian of gravity theories whose action is linear in spacetime curvature}

\author{Yuri Bonder}

\address{Instituto de Ciencias Nucleares, Universidad Nacional Aut\'onoma de M\'exico\\
Apartado Postal 70-543, Coyoac\'an 04510 Cd.Mx., M\'exico\\bonder@nucleares.unam.mx
}

\begin{abstract}
A straightforward method to compute Hamilton's density for theories that are linear in the spacetime curvature is provided. It is shown that the lapse function and shift vector still give rise to primary constraints, while the induced metric gives rise to nontrivial evolution equations. The corresponding Hamilton's density can always be obtained, albeit in a formal sense.
\end{abstract}

\bodymatter

Many aspects of GR were clarified when studied from a hamiltonian viewpoint. This include, for example, the number and classification of constraints, the well-known ADM mass, and some notions of stability. On the other hand, people have consider extensions to general relativity (GR). It can thus be expected that analogous clarifications would become available for these new theories.

Among those alternative theories of gravity, it is interesting to consider those that break local Lorentz invariance. This is mainly because local Lorentz invariance is one of the most basic principles of modern physics. The most natural way to break such a principle is with a generic parametrization, as done within the Standard-Model Extension\cite{SME} (SME). It should be mentioned that the hamiltonian framework is natural for some Lorentz violating theories as it is associated with a foliation and the algebra of constraints has valuable information on diffeomorphisms invariance, which is affected when Lorentz is explicitly violated\cite{nosotros}. In addition, it allows one to make a rigorous counting of degrees of freedom and to find all the constraints, which, in turn, are necessary to perform numerical analyses.

A method to study the Hamiltonian formulation for theories whose action is linear in the spacetime curvature is presented. This includes GR and the minimal sector of the gravitational SME. This contribution should be regarded as a generalization to existing work\cite{SMEHam} in that it is exact (no truncation on the Lorentz violating parameters, known as SME coefficients, is required), it is general (no conditions on these coefficients are assumed), and it is fully covariant.

Regarding the notation: abstract indexes are represented by Latin characters from the beginning of the alphabet; they should be thought of as slots in the conventional tensorial notation, and thus, no coordinates are required. Index contraction is represented by repeating an index. A pair of indexes in parenthesis (brackets) represent its symmetric (antisymmetric) part, with a $1/2$ factor. Importantly, no notational distinction is made between spacetime tensors and tensors associated with a submanifold. Spacetime is assumed to be a $4$-dimensional manifold with a Lorentzian metric $g_{ab}$; $g_{ab}$ and its inverse are used to lower and raise abstract indexes.

Clearly, spacetime must be globally hyperbolic\cite{Wald} so that it is foliated by Cauchy surfaces $\Sigma_t$. To proceed, a vector field $t^{a}$ whose integral lines can be used to identify points in different Cauchy surfaces must be provided. This vector can be decomposed as $t^{a}=Nn^a +N^a$, where $N>0$ is the lapse function and $N^a$ is the shift vector (it is a vector in $\Sigma_t$). Let $n^a$ be the unit normal vector to $\Sigma_t$, then, the induced metric on $\Sigma_t$ becomes $h_{ab}= g_{ab}+ n_a n_b$. What is more, $h_a^b$ is the projector into $\Sigma_t$. Also, it is possible to define the extrinsic curvature of $\Sigma_t$, $k_{ab}$, which describes how $\Sigma_t$ bends in spacetime. There are several expressions for $k_{ab}$, the most relevant for this contribution is $k_{ab} = h^c_a \nabla_c n_b$. In addition, it is possible to define the time derivative associated to $t^a$, which acts on any tensor and is denoted by a dot. This derivative is obtained by taking the Lie derivative with respect to $t^a$ and projecting all the indexes into $\Sigma_t$. For example,
\begin{eqnarray}
\dot{h}_{ab}=2Nk_{ab}+2D_{(a}N_{b)},\label{doth}
\end{eqnarray}
where $D_a$ is the derivative operator in $\Sigma_t$ associated with $h_{ab}$.

This contribution studies metric theories of gravity whose action is linear in the spacetime curvature. Clearly, this curvature tensor must be contracted with an indexed object that may be a function of the metric and/or nondynamical objects like the SME coefficients. Also, for simplicity, no matter fields are considered and all boundary terms are ignored. Using the well-known Gauss-Codazzi relations\cite{Wald}, it can be shown that any action in this category, when written in terms of $h_{ab}$, $k_{ab}$, $N$, and $N^a$, can be brought to the form
\begin{equation}
S=\int d^4 x N \sqrt{h}\left(k_{ab}A^{abcd}k_{cd}+ 2k_{ab}B^{ab}+C \right),
\end{equation}
where $A^{abcd}=A^{(ab)(cd)}=A^{cdab}=A^{abcd}(h)$, $B^{ab}=B^{ba}=B^{ab}(N,\vec{N},h)$, $C=C(h)$. Also, $h$ in the action is the determinant of the components of $h_{ab}$. Note that, besides the global factor $N$, which is associated to the spacetime volume element, $N$ and $N^a$ only appear within $B^{ab}$. It is also important to notice that there are no time derivatives of $N$ and $N^a$, implying that their conjugate momenta vanishes, and thus, that the action variations with respect to $N$ and $N^a$ are primary constraints. In these theories the evolution is associated to the fact that $\dot{h}_{ab}$ appears in $k_{ab}$, as can be seen in Eq.~\eqref{doth}. Therefore, the conjugate momenta to $h_{ab}$ takes the form
 \begin{equation}
\pi^{ab}=\frac{\delta S}{\delta \dot{h}_{ab}}=\sqrt{h}\left(A^{abcd}k_{cd}+ B^{ab} \right).\label{pi}
\end{equation}

To write the corresponding Hamilton density, Eq.~\eqref{pi} must be inverted. That is, is it necessary to find $\alpha_{abcd}=\alpha_{(ab)(cd)}=\alpha_{abcd}(h)$ such that $\alpha_{abcd}A^{cdef}=h^e_{(a} h^f_{b)}$. Provided that $\alpha_{abcd}$ can be found, it is possible to write $k_{cd}= \alpha_{cdab}p^{ab}$, where $p^{ab}\equiv\pi^{ab}/\sqrt{h}-B^{ab}$. Hence, Hamilton's density takes the very compact form
\begin{equation}
\mathcal{H}= 2\pi^{ab}D_{a}N_{b} + N\sqrt{h}p^{ab}\alpha_{abcd}p^{cd}- N\sqrt{h}C.
\end{equation}
To find an expression for $\alpha_{abcd}$, it is useful to assume that $A^{abcd}$ has a piece that is proportional to $h^{ac} h^{db}$, namely, $A^{abcd}=a h^{ac} h^{db} -\delta a^{abcd}$, where $a$ and $\delta a^{abcd} $ may depend on $h_{ab}$ and the SME coefficients. Assuming $a\neq 0$, it is possible to write a purely formal expression for $\alpha_{abcd} $:
\begin{equation}
\alpha_{abcd}  = \frac{1}{a}  \left[ h_{a(c}h_{d)b}-\frac{1}{2}h_{ab}h_{cd}+  \left(h^e_{c} h^f_{d}-\frac{1}{2}h_{cd}h^{ef}\right)\sum_{n=1}^\infty \frac{c^{(n)}_{abef}}{a^n}\right] ,\label{alpha}
\end{equation}
where the $c^{(n)}_{abcd}$ are defined by the recurrence relation $c^{(n)}_{abcd} \delta a^{cdef}={c^{(n+1)}}_{ab}\ ^{ef}$ for $n\geq1$, with $ c^{(1)}_{abcd}=\delta a_{abcd}$ as a seed.

The evolution equations take the form
\begin{eqnarray}
\dot{h}_{ab}&=&2 D_{(a}N_{b)} + 2N\alpha_{abcd}p^{cd} ,\label{hdot1}\\
\dot{\pi}_{ab}&=&-2\pi^{c(a} D_c N^{b)} + \sqrt{h}D_c\left(\frac{N^c \pi^{ab}}{ \sqrt{h}}\right) + \sqrt{h}Nh^{ab} p^{cd}\alpha_{cdef}\left(\frac{p^{ef}}{2}+B^{ef}\right)\nonumber\\
&& + \frac{N\sqrt{h}}{2}h^{ab}C -  \int d^3 x \sqrt{h}Np^{cd} \frac{\delta\alpha_{cdef}}{\delta h_{ab}}p^{ef}+ \int d^3x  \sqrt{h}N\frac{\delta C}{\delta h_{ab}}\nonumber\\
&&+ 2 \int d^3 x \sqrt{h}Np^{cd}\alpha_{cdef} \frac{\delta B^{ef}}{\delta h_{ab}}. \label{pidot}
\end{eqnarray}
Note that Eq.~\eqref{hdot1} is equivalent to Eq.~\eqref{doth}. Now, the primary constraints are
\begin{eqnarray}
0&=& \sqrt{h}p^{ab}\alpha_{abcd}p^{cd}- \sqrt{h}C- 2\int d^3 x \sqrt{h}Np^{ab}\alpha_{abcd}\frac{\delta B^{cd}}{\delta N},\\
0&=&-2\sqrt{h}D_b\left(\frac{{\pi_a}^{b}}{\sqrt{h}}\right)- 2\int d^3 x \sqrt{h}N p^{cd}\alpha_{cdef} \frac{\delta B^{ef}}{\delta N^a}.\end{eqnarray}
This is not enough to have a well posed evolution; one needs to verify if there are additional expressions needed to guarantee that the constraints are kept valid during the evolution. In addition, one should classify the constraints as first and second class constraints\cite{Dirac}. This can be done by obtaining the algebra of the constraints, that is, the Poisson brackets among all the (smeared) constraints. However, to find this algebra, a concrete theory must be given, and still, the computations are extremely challenging. Clearly, the existence of nontrivial constraints is related to the fact that, for explicit Lorentz violation, the SME coefficients are not generic\cite{explicit}. It is interesting that, at this point, it is possible to verify that $B^{ab}$ governs the nature of the constraint algebra, since, for example, in the case where $\delta B^{ab}/\delta N=0=\delta B^{ab}/\delta N^c$, Hamilton's density vanishes (on shell), which is closely related with invariance under diffeomorphisms. Finally, this formalism must generate expressions that are equivalent to those obtained in a lagrangian formulation. Nevertheless, showing such an equivalence is also nontrivial and it often relies on knowing all the constraints.

\section*{Acknowledgments}
I benefited from interactions with C.\ Peterson, C.M.\ Reyes, and M.\ Schreck. This research was financially supported by UNAM-DGAPA-PAPIIT Grant IG100120 and CONACyT FORDECYT-PRONACES grant 140630.

\end{document}